\documentclass{PoS}
\let\ifpdf\relax
\usepackage{natbib,colortbl,multirow,graphicx,epstopdf,graphics,amsmath}
\usepackage{xcolor}

\definecolor{lightgray}{gray}{0.9}
\definecolor{lightblue}{rgb}{0.11,0.56,1}
\definecolor{mag}{rgb}{1,0,1}
\definecolor{cyan}{rgb}{0.88,1,1}
\definecolor{green}{rgb}{0.49,0.99,0}
\definecolor{yellow}{rgb}{1,1,0}

\title{Weak Lensing Simulations for the SKA}

\ShortTitle{SKA Weak Lensing Simulations}

\author{\speaker{Prina Patel}$^{a,b}$, Ian Harrison$^{c}$, Sphesihle Makhathini$^{d,b}$, Filipe Abdalla$^{e,d}$, David Bacon$^{f}$, Michael L. Brown$^{c}$, Ian Heywood$^{g,d}$, Matt Jarvis$^{h,a}$, Oleg Smirnov$^{d,b}$\\
$^{a}$University of the Western Cape,
$^{b}$SKA South Africa,
$^{c}$Jodrell Bank,
$^{d}$Rhodes University,
$^{e}$University College London,
$^{f}$ICG,
$^{g}$CSIRO,
$^{h}$University of Oxford,

        E-mail: \email{prina83@gmail.com}}

\abstract{Weak gravitational lensing is a very promising probe for cosmology. Measurements are traditionally made at optical wavelengths where many highly resolved galaxy images are readily available. However, the Square Kilometre Array (SKA) holds great promise for this type of measurement at radio wavelengths owing to its greatly increased sensitivity and resolution over typical radio surveys. The key to successful weak lensing experiments is in measuring the shapes of detected sources to high accuracy. In this document we describe a simulation pipeline designed to simulate radio images of the quality required for weak lensing, and will be typical of SKA observations. We provide as input, images with realistic galaxy shapes which are then simulated to produce images as they would have been observed with a given radio interferometer. We exploit this pipeline to investigate various stages of a weak lensing experiment in order to better understand the effects that may impact shape measurement. We first show how the proposed SKA1-Mid array configurations perform when we compare the (known) input and output ellipticities. We then investigate how making small changes to these array configurations impact on this input-outut ellipticity comparison. We also demonstrate how alternative configurations for SKA1-Mid that are smaller in extent, and with a faster survey speeds produce similar performance to those originally proposed. We then show how a notional SKA configuration performs in the same shape measurement challenge. Finally, we describe ongoing efforts to utilise our simulation pipeline to address questions relating to how applicable current (mostly originating from optical data) shape measurement techniques are to future radio surveys. As an alternative to such image plane techniques, we lastly discuss a shape measurement technique based on the shapelets formalism that reconstructs the source shapes directly from the visibility data. We end with a discussion of extensions to the out current simulations and concluding remarks. }

\FullConference{
Advancing Astrophysics with the Square Kilometre Array\\
June 8-13, 2014\\
Giardini Naxos, Italy}

\begin{document}

\section{Introduction}
Weak gravitational lensing is a promising probe for cosmology. Light rays from distant sources are bent by the gravitational potential of objects on the path to an observer, leading to a coherent ellipticity or shear on images of galaxies near each other on the sky. We can measure statistics of this shear for galaxies in the Universe, at different redshifts and angular separations. These statistics are sensitive to the growth history of density fluctuations in the Universe (and therefore the matter power spectrum), and to the expansion history of the Universe (and hence for instance, dark energy parameters).

Weak shear measurements are already maturing at optical frequencies \citep[e.g.][]{2013MNRAS.430.2200K}, and a range of future optical experiments are planned to provide tight constraints on cosmological parameters using this probe (for instance the ground based Large Synoptic Survey Telescope (LSST), \citep{2009arXiv0912.0201L}, and Euclid space telescope, \citep{2012SPIE.8442E..0TL}. See also \citet{LSSTSKASCIENCEBOOK} and \citet{EUCLIDSKASCIENCEBOOK}). In addition, shear can be measured at radio frequencies \citep[e.g.][]{2004ApJ...617..794C, 2010MNRAS.401.2572P}, and as shown in \citet{WLSKASCIENCEBOOK} SKA will be able to provide competitive gravitational lensing measurements.

In order to demonstrate this, it is necessary to simulate realistic images which could be obtained using particular SKA measurement sets, including the effects of realistic gravitational lensing and telescope effects. Shear measurement techniques should be carried out on these images, and we should verify that we can obtain faithful estimates of the true shear. In this chapter, we describe our efforts to make such simulations and confirm that SKA configurations which are being considered by the community are suitable for making the SKA a powerful gravitational lensing telescope.

\section{Simulation Pipeline}
The simulations we utilise are based on a pipeline built in two parts which we briefly describe here and refer to \citet{2013arXiv1303.4650P} for further details. The pipeline works by taking input images that contain realistic galaxy shapes and running them through a simulator. Within the simulator we can define the interferometer to use and other observation details such as the frequency, bandwidth, integration time etc. The simulator then predicts the visibilities for the given image and observation. The last part of the simulation then takes these visibilities and images them as would be done with real observed data and a restored image is produced. Once the simulation has produced a final restored image we then re-analyse these images with our chosen shape measurement technique and compare the input and out ellipticities.

\subsection{Simulation}
We firstly describe the input images we have created that are used throughout this work. The input images are based on a shapelet (see \S \ref{sec:shapelets}) based method that is described in detail in \citet{2003MNRAS.338...35R} and \citet{2003MNRAS.338...48R}. Briefly, the shapelets method decomposes a galaxy image into a series of localised basis functions called shapelets. The shapelets are a complete and orthonormal set of basis functions consisting of weighted Hermite polynomials, corresponding to perturbations around a circular Gaussian. The process of generating these input images is described in detail in \citep{2013MNRAS.435..822R}, resulting in shapelet models that represent simulated, but realistic, (known) galaxy shapes as would have been observed with the Hubble Space Telescope (HST). Although these are simulated optical galaxies we make use of them as there exists no large enough sample of highly resolved galaxy images in the radio. In \citet{2010MNRAS.401.2572P} it was found that on a case-by-case basis the intrinsic shapes of radio and optical sources were only weakly  correlated, but that the overall distribution of ellipticities were very similar at the two wavelengths. To keep computation time low we have created 100 such images that are $0.85\times0.85$ arcminutes$^{2}$ with a pixel scale of $0.05$ arc seconds, that each contain $\simeq100$ sources each. This gives a resulting input number density of $n\simeq140$ arcminutes$^{-2}$, which is far larger than any current lensing survey. 

These images are then fed into the simulator which in combination with the chosen interferometer and observation particulars, predicts the visibilities. If desired we are then able to include affects that would effect the visibilities in a variety of ways using the Radio Interferometer Measurement Equation (RIME) formalism \citep[e.g.][]{2011A&A...527A.106S}, which relates the propagation of the signal from the source to detector via various observational effects. We are also able to include Gaussian measurement noise on the visibilities. Note in our simulations we do not currently employ any observation effects but we do include Gaussian measurement noise. These visibilities are then imaged using standard techniques (e.g. CLEAN) that we are again able to control, generating the output restored image. 

\subsection{Shape Measurement}
We take the restored images produced by the simulation and then analyse them using the image based shapelet method described in \S\ref{sec:shapelets} and compare the input and output ellipticities. Note, that in producing the restored image using CLEAN, a convolution is performed between the model image and the main lobe of the synthesised beam (PSF). We simulate this PSF as well and perform a deconvolution within the shapelet analysis, further details on deconvolution with shapelets can be found in \citet{2013MNRAS.435..822R}, and further details about the image plane shapelet analysis can be found in \citet{2013arXiv1303.4650P}. We are then left with shapelet models (i.e. shapelet coefficients $f_{n,m}$) both before simulation and post, from which we estimate a 2-component (complex) ellipticity according to
\begin{equation}
\epsilon=\frac{\sqrt{2}f^{\prime}_{2,2}}{\langle f_{0,0}-f_{4,0}\rangle},
\end{equation}
\citep{2007MNRAS.380..229M}. This ellipticity estimator is the Gaussian weighted quadrupole moment cast into shapelet space. We then follow \citet{2006MNRAS.368.1323H} and fit a linear model to our data points
\begin{equation}
\epsilon_{i}-\epsilon_{i}^{t}=m_{i}\epsilon_{i}^{t}+c_{i},
\end{equation}
where $\epsilon_{i}^{t}$ is the true input ellipticity and $\epsilon_{i}$ is the measured ellipticity. In all the follows the relative merit of each experiment can be compared through the calculated values of $m_{i}$ and $c_{i}$. For a perfect experiment where we fully recover exactly all the shapes in the images $m_{i}=0, c_{i}=0$. A non-zero $m_{i}$ is indicative of a calibration bias resulting from poor correction of factors that circularise images, poor PSF correction for example. $c_{i}\neq0$ suggests a systematic where even circular objects appear elliptical. We have made no attempt to optimise any of our analyses to reduce the biases in anyway, neither have we looked for the origin of these biases. As such, the bias values presented here are only meaningful in a comparative sense and should not be taken to represent the final performance that any such experiment might achieve. In a full analysis one would hope to understand the nature and origin of such biases to high precision in order to correct for them. One clear use of our pipeline is to asses the levels of bias that may be introduced by observation effects (e.g. Direction Dependant Effects (DDEs)). 

The key to making a weak lensing measurement requires accurate measurements of many galaxy shapes. Our simulation pipeline offers a way in which all parts of the data processing pipeline, from raw visibilities to restored images, can be explored. In the rest of this document we present investigations that have been carried out to address some of the key questions most relevant to weak lensing studies with the SKA. 

\section{SKA Baselines Configurations}
\label{sec:configs}
Since the configuration for SKA1-Mid is yet to be completely finalised we explore how changing the array configuration as proposed by a small amount would effect the ability to accurately measure the shapes of galaxies. We generated many different SKA1-Mid array configurations and ran them through our pipeline. In this section we describe what impact minor changes had on the calculated calibration and additive biases. 
\subsection{Baselines Changes and Impact on Weak Lensing}
\label{sec:ska1configs}
\begin{figure*}
\centering
\includegraphics[scale=0.4]{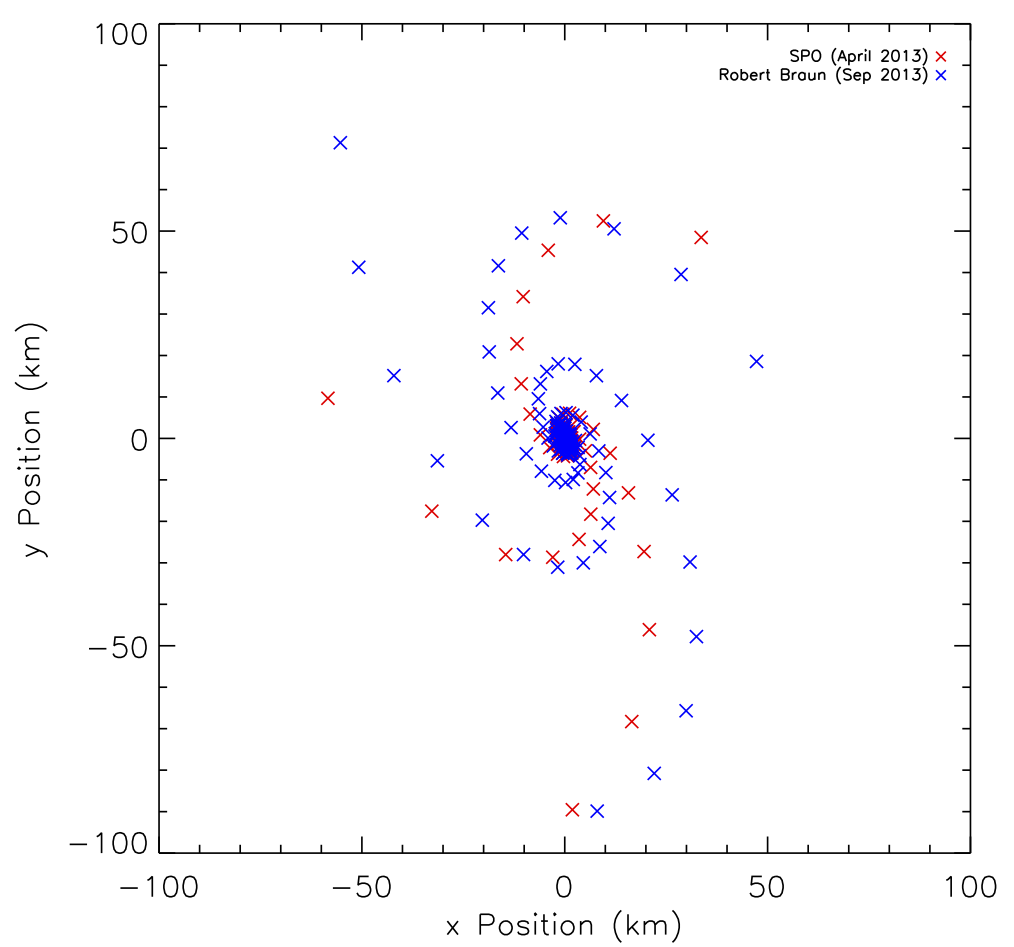}  \includegraphics[scale=0.40]{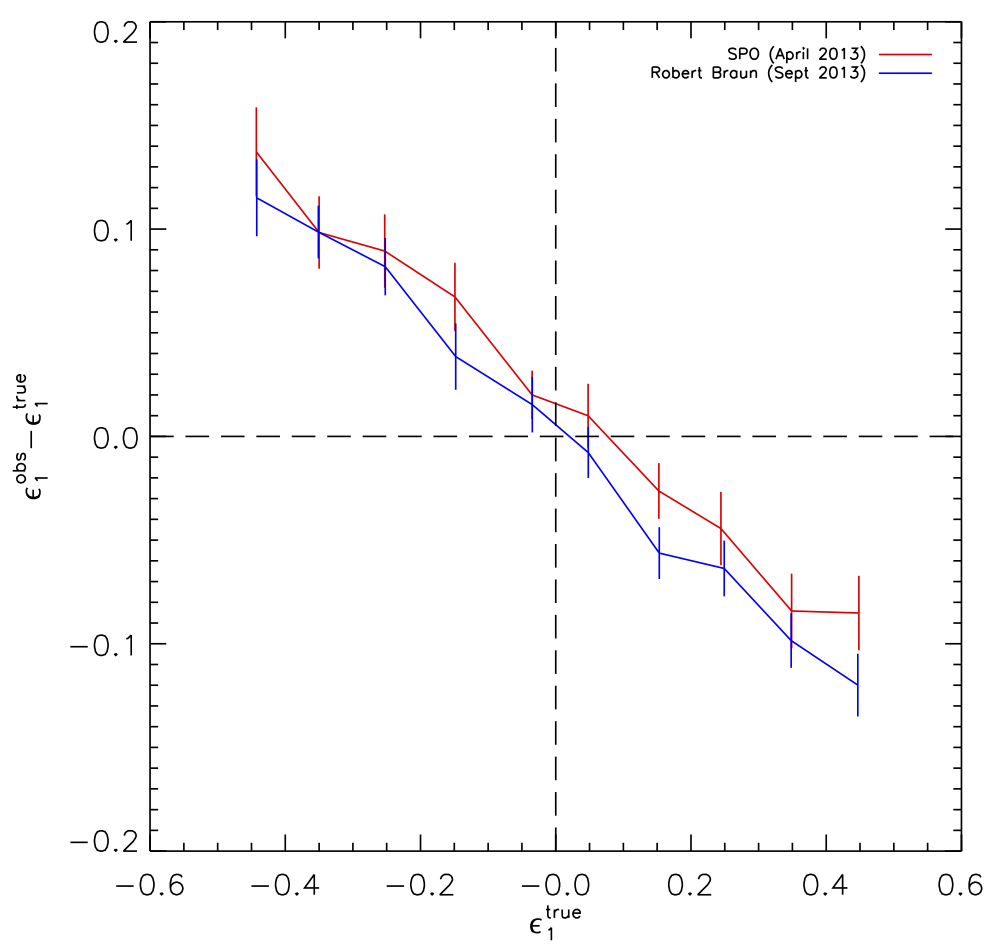}
\caption{{\it Left panel:} Array configurations for SKA1-Mid as proposed by the SKAO (April 2013) and Robert Braun (September 2013). {\it Right Panel:} Recovered ellipticity (only component 1 of the ellipticity is shown) distributions for these SKA1-Mid arrays. \label{fig:prop3ref2configeineout}}
\end{figure*}

We initially calculated the bias values for the two proposed SKA1-Mid configurations. The first is that proposed by the SKAO (referred to as SKAM) and the other, a short time after, by Robert Braun (SKAM12), both are shown in Figure \ref{fig:prop3ref2configeineout}. Both these arrays contain 254 dishes with 197 of them within a 4 km core, and the other 57 divided into 3 logarithmically spaced spiral arms extending out to 100 km. For all simulations run in this section we have adopted an 8 hour observation at 700MHz and 10 50 MHz channels pointing at declination $\delta=-40^{\circ}$. We add Gaussian measurement noise to the visibilities resulting in the sources in the output images having a signal-to-noise of $\simeq10$. We have also adopted a uniform weighting scheme through this work. Also shown in Figure \ref{fig:prop3ref2configeineout} are the recovered ellipticity distributions derived for both configurations. Unsurprisingly, both these two configurations produce similar calibration values, with $m_{i}\simeq-0.261$ and very small additive bias. We use these recovered values of $m_{i}$ and $c_{i}$ as our base values to which we can compare the values derived from modified SKA1-Mid configurations. Note that all the calculated bias values for all considered arrays are given in Table \ref{tab:configresults}.

The changes we explored were: changing the spacing in the arms, taking dishes from the core and redistributing them into the 3 spiral arms and adding new dishes to the spiral arms. In the former case, we looked at changing the arm distribution from logarithmic to equidistant and linear. In the case of the latter two, 9, 21, 30, 39, 51, and 60 dishes were added to the arms while keeping the maximum extent of the arms the same. In Table \ref{tab:configresults} we show the ellipticity recovery performance of these other SKA1-Mid configurations described above. The first two entries in blue and magenta correspond to the SKAO and Robert Braun configurations respectively. The cyan rows correspond to the configurations with equidistant and linear arms spacing. Entries in green and yellow are those where either dishes were redistributed from the core (green), or new dishes (yellow) were added to the spiral arms.  


\begin{table}
\caption{SKA1-Mid ellipticity recovery performance results. The first two entries in blue and magenta correspond to the SKAO and Robert Braun configurations respectively. The cyan rows correspond to the configurations with equidistant and linear arms spacing. Entries in green and yellow are those where either dishes were redistributed from the core (green), or new dishes (yellow) were added to the spiral arms. 
${}^{a}$~quoted relative to the SKAO and Robert Braun configurations.\label{tab:configresults}  }
\centering
\resizebox{\linewidth}{!}{
\begin{tabular}{|c|c|c|c|c|c|c|c|c|}
\hline
Name & $N_{total}$ & $N_{core}$ & $N_{arms}$ & Arm & $^{a}$Sensitivity for & Number Density & $m_{i}$ & $c_{i}$ \\
& Dishes & Dishes & Dishes & Spacing & SNR $=10$ & $n$ arcminute$^{-2}$\\
\hline
\hline
\cellcolor{lightblue}  & \cellcolor{lightblue} & \cellcolor{lightblue} & \cellcolor{lightblue} & \cellcolor{lightblue} &\cellcolor{lightblue} & 
\cellcolor{lightblue} & \cellcolor{lightblue} $-0.278\pm0.021$ & \cellcolor{lightblue}$0.001\pm0.005$  \\
\multirow{-2}{*}{\cellcolor{lightblue}SKAM (SKAO)} & \multirow{-2}{*}{\cellcolor{lightblue}254} & \multirow{-2}{*}{\cellcolor{lightblue}197} & \multirow{-2}{*}{\cellcolor{lightblue}19} & 
\multirow{-2}{*}{\cellcolor{lightblue}Logarithmic} & \multirow{-2}{*}{\cellcolor{lightblue}1.0} &  \multirow{-2}{*}{\cellcolor{lightblue}21.97} & \cellcolor{lightblue}$-0.258\pm0.020$ & \cellcolor{lightblue}$0.017\pm0.005$ \\
\hline
\cellcolor{mag}& \cellcolor{mag}&\cellcolor{mag} & \cellcolor{mag}&\cellcolor{mag} &\cellcolor{mag} & \cellcolor{mag} & $\cellcolor{mag}-0.227\pm0.015$ & $\cellcolor{mag}-5\times 10^{-4}\pm0.004$ \\
\multirow{-2}{*}{\cellcolor{mag}SKAM12 (Robert Braun)} & \multirow{-2}{*}{\cellcolor{mag}254} & \multirow{-2}{*}{\cellcolor{mag}197} & \multirow{-2}{*}{\cellcolor{mag}19} & \multirow{-2}{*}{\cellcolor{mag}Logarithmic} & \multirow{-2}{*}{\cellcolor{mag}1.0} & \multirow{-2}{*}{\cellcolor{mag}30.43} & $\cellcolor{mag}-0.280\pm0.016$ & $\cellcolor{mag}0.001\pm0.004$ \\ 
\hline
\cellcolor{cyan}& \cellcolor{cyan}& \cellcolor{cyan}&\cellcolor{cyan}& \cellcolor{cyan}& \cellcolor{cyan}&  \cellcolor{cyan}&$\cellcolor{cyan}-0.319\pm0.015$ & $\cellcolor{cyan}-0.006\pm0.004$ \\
\multirow{-2}{*}{\cellcolor{cyan}SKAM12EQ} & \multirow{-2}{*}{\cellcolor{cyan}254} & \multirow{-2}{*}{\cellcolor{cyan}197} & \multirow{-2}{*}{\cellcolor{cyan}19} & \multirow{-2}{*}{\cellcolor{cyan}Equidistant} & \multirow{-2}{*}{\cellcolor{cyan}1.25} &  \multirow{-2}{*}{\cellcolor{cyan}19.32}& $\cellcolor{cyan}-0.301\pm0.015$ & $\cellcolor{cyan}0.010\pm0.004$ \\ 
\hline
\cellcolor{cyan}&\cellcolor{cyan} &\cellcolor{cyan} &\cellcolor{cyan} &\cellcolor{cyan} & \cellcolor{cyan}& \cellcolor{cyan} &$\cellcolor{cyan}-0.297\pm0.015$ & $\cellcolor{cyan}-0.006\pm0.004$ \\
\multirow{-2}{*}{\cellcolor{cyan}SKAM12LIN} & \multirow{-2}{*}{\cellcolor{cyan}254} & \multirow{-2}{*}{\cellcolor{cyan}197} & \multirow{-2}{*}{\cellcolor{cyan}19} & \multirow{-2}{*}{\cellcolor{cyan}Linear} & \multirow{-2}{*}{\cellcolor{cyan}1.0} & \multirow{-2}{*}{\cellcolor{cyan}30.69} &$\cellcolor{cyan}-0.282\pm0.016$ & $\cellcolor{cyan}0.004\pm0.004$ \\ 
\hline
\cellcolor{green}&\cellcolor{green} & \cellcolor{green}&\cellcolor{green} &\cellcolor{green} & \cellcolor{green}&\cellcolor{green} & $\cellcolor{green}-0.292\pm0.019$ & $\cellcolor{green}-0.006\pm0.005$ \\
\multirow{-2}{*}{\cellcolor{green}SKAM9C} & \multirow{-2}{*}{\cellcolor{green}254} & \multirow{-2}{*}{\cellcolor{green}188} & \multirow{-2}{*}{\cellcolor{green}22} & \multirow{-2}{*}{\cellcolor{green}Logarithmic} & \multirow{-2}{*}{\cellcolor{green}1.0} & \multirow{-2}{*}{\cellcolor{green}32.20} & $\cellcolor{green}-0.264\pm0.017$ & $\cellcolor{green}0.002\pm0.005$ \\ 
\hline
\cellcolor{green}& \cellcolor{green}&\cellcolor{green} & \cellcolor{green}& \cellcolor{green}&\cellcolor{green} &\cellcolor{green} & $-\cellcolor{green}0.307\pm0.018$ & $\cellcolor{green}-0.002\pm0.005$ \\
\multirow{-2}{*}{\cellcolor{green}SKAM21C} & \multirow{-2}{*}{\cellcolor{green}254} & \multirow{-2}{*}{\cellcolor{green}176} & \multirow{-2}{*}{\cellcolor{green}26} & \multirow{-2}{*}{\cellcolor{green}Logarithmic} & \multirow{-2}{*}{\cellcolor{green}1.0} & \multirow{-2}{*}{\cellcolor{green}32.33} & $\cellcolor{green}-0.278\pm0.019$ & $\cellcolor{green}0.001\pm0.005$ \\ 
\hline
\cellcolor{green}&\cellcolor{green} & \cellcolor{green}& \cellcolor{green}& \cellcolor{green}&\cellcolor{green} &\cellcolor{green} & $\cellcolor{green}-0.308\pm0.015$ & $\cellcolor{green}-0.006\pm0.004$ \\
\multirow{-2}{*}{\cellcolor{green}SKAM30C} & \multirow{-2}{*}{\cellcolor{green}254} & \multirow{-2}{*}{\cellcolor{green}167} & \multirow{-2}{*}{\cellcolor{green}29} & \multirow{-2}{*}{\cellcolor{green}Logarithmic} & \multirow{-2}{*}{\cellcolor{green}1.0} & \multirow{-2}{*}{\cellcolor{green}30.67} & $\cellcolor{green}-0.286\pm0.017$ & $\cellcolor{green}0.001\pm0.005$ \\ 
\hline
\cellcolor{green}& \cellcolor{green}& \cellcolor{green}&\cellcolor{green} & \cellcolor{green}&\cellcolor{green} &\cellcolor{green} & $\cellcolor{green}-0.333\pm0.014$ & $\cellcolor{green}-0.006\pm0.004$ \\
\multirow{-2}{*}{\cellcolor{green}SKAM39C} & \multirow{-2}{*}{\cellcolor{green}254} & \multirow{-2}{*}{\cellcolor{green}158} & \multirow{-2}{*}{\cellcolor{green}32} & \multirow{-2}{*}{\cellcolor{green}Logarithmic} & \multirow{-2}{*}{\cellcolor{green}1.0} & \multirow{-2}{*}{\cellcolor{green}25.41} & $\cellcolor{green}-0.293\pm0.015$ & $\cellcolor{green}0.011\pm0.004$ \\ 
\hline
\cellcolor{green}&\cellcolor{green} &\cellcolor{green} &\cellcolor{green} &\cellcolor{green} &\cellcolor{green} &\cellcolor{green} & $\cellcolor{green}-0.314\pm0.016$ & $\cellcolor{green}-0.005\pm0.005$ \\
\multirow{-2}{*}{\cellcolor{green}SKAM51C} & \multirow{-2}{*}{\cellcolor{green}254} & \multirow{-2}{*}{\cellcolor{green}146} & \multirow{-2}{*}{\cellcolor{green}36} & \multirow{-2}{*}{\cellcolor{green}Logarithmic} & \multirow{-2}{*}{\cellcolor{green}1.13}& \multirow{-2}{*}{\cellcolor{green}31.21}  & $\cellcolor{green}-0.285\pm0.015$ & $\cellcolor{green}0.008\pm0.004$ \\ 
\hline
\cellcolor{green}&\cellcolor{green} &\cellcolor{green} &\cellcolor{green} & \cellcolor{green}&\cellcolor{green} &\cellcolor{green} & $\cellcolor{green}-0.318\pm0.015$ & $\cellcolor{green}-0.005\pm0.004$ \\
\multirow{-2}{*}{\cellcolor{green}SKAM60C} & \multirow{-2}{*}{\cellcolor{green}254} & \multirow{-2}{*}{\cellcolor{green}137} & \multirow{-2}{*}{\cellcolor{green}39} & \multirow{-2}{*}{\cellcolor{green}Logarithmic} & \multirow{-2}{*}{\cellcolor{green}1.13} &  \multirow{-2}{*}{\cellcolor{green}25.72} &$\cellcolor{green}-0.297\pm0.015$ & $\cellcolor{green}0.005\pm0.004$ \\ 
\hline
\cellcolor{yellow}&\cellcolor{yellow} & \cellcolor{yellow}& \cellcolor{yellow}&\cellcolor{yellow} & \cellcolor{yellow}& \cellcolor{yellow} & $\cellcolor{yellow}-0.234\pm0.016$ & $\cellcolor{yellow}-0.007\pm0.004$ \\
\multirow{-2}{*}{\cellcolor{yellow}SKAM263} & \multirow{-2}{*}{\cellcolor{yellow}263} & \multirow{-2}{*}{\cellcolor{yellow}197} & \multirow{-2}{*}{\cellcolor{yellow}22} & \multirow{-2}{*}{\cellcolor{yellow}Logarithmic} & \multirow{-2}{*}{\cellcolor{yellow}1.13} &\multirow{-2}{*}{\cellcolor{yellow}33.52} & $\cellcolor{yellow}-0.264\pm0.016$ & $\cellcolor{yellow}-0.003\pm0.004$ \\ 
\hline
\cellcolor{yellow}&\cellcolor{yellow} &\cellcolor{yellow} &\cellcolor{yellow} &\cellcolor{yellow} &\cellcolor{yellow} &\cellcolor{yellow} &  $\cellcolor{yellow}-0.256\pm0.017$ & $\cellcolor{yellow}-0.004\pm0.005$ \\
\multirow{-2}{*}{\cellcolor{yellow}SKAM275} & \multirow{-2}{*}{\cellcolor{yellow}275} & \multirow{-2}{*}{\cellcolor{yellow}197} & \multirow{-2}{*}{\cellcolor{yellow}26} & \multirow{-2}{*}{\cellcolor{yellow}Logarithmic} & \multirow{-2}{*}{\cellcolor{yellow}1.13} &\multirow{-2}{*}{\cellcolor{yellow}29.61} & $\cellcolor{yellow}-0.256\pm0.018$ & $\cellcolor{yellow}-0.009\pm0.005$ \\ 
\hline
\cellcolor{yellow} & \cellcolor{yellow} &\cellcolor{yellow} &\cellcolor{yellow} &\cellcolor{yellow} &\cellcolor{yellow} &\cellcolor{yellow} &  $\cellcolor{yellow}-0.256\pm0.017$ & $\cellcolor{yellow}-0.002\pm0.005$ \\
\multirow{-2}{*}{\cellcolor{yellow}SKAM284} & \multirow{-2}{*}{\cellcolor{yellow}284} & \multirow{-2}{*}{\cellcolor{yellow}197} & \multirow{-2}{*}{\cellcolor{yellow}29} & \multirow{-2}{*}{\cellcolor{yellow}Logarithmic} & \multirow{-2}{*}{\cellcolor{yellow}1.13} &\multirow{-2}{*}{\cellcolor{yellow}29.79} & $\cellcolor{yellow}-0.268\pm0.017$ & $\cellcolor{yellow}0.002\pm0.005$ \\ 
\hline
\cellcolor{yellow}&\cellcolor{yellow} &\cellcolor{yellow} &\cellcolor{yellow} &\cellcolor{yellow} &\cellcolor{yellow} &\cellcolor{yellow} &  $\cellcolor{yellow}-0.260\pm0.018$ & $\cellcolor{yellow}-0.003\pm0.005$ \\
\multirow{-2}{*}{\cellcolor{yellow}SKAM293} & \multirow{-2}{*}{\cellcolor{yellow}293} & \multirow{-2}{*}{\cellcolor{yellow}197} & \multirow{-2}{*}{\cellcolor{yellow}32} & \multirow{-2}{*}{\cellcolor{yellow}Logarithmic} & \multirow{-2}{*}{\cellcolor{yellow}1.25} & \multirow{-2}{*}{\cellcolor{yellow}29.87} &$\cellcolor{yellow}-0.269\pm0.016$ & $\cellcolor{yellow}-0.001\pm0.004$ \\ 
\hline
\cellcolor{yellow}&\cellcolor{yellow} &\cellcolor{yellow} &\cellcolor{yellow} &\cellcolor{yellow} &\cellcolor{yellow} &\cellcolor{yellow} &  $\cellcolor{yellow}-0.272\pm0.016$ & $\cellcolor{yellow}-0.005\pm0.004$ \\
\multirow{-2}{*}{\cellcolor{yellow}SKAM305} & \multirow{-2}{*}{\cellcolor{yellow}305} & \multirow{-2}{*}{\cellcolor{yellow}197} & \multirow{-2}{*}{\cellcolor{yellow}36} & \multirow{-2}{*}{\cellcolor{yellow}Logarithmic} & \multirow{-2}{*}{\cellcolor{yellow}1.25} & \multirow{-2}{*}{\cellcolor{yellow}30.09} &$\cellcolor{yellow}-0.281\pm0.016$ & $\cellcolor{yellow}0.002\pm0.005$ \\ 
\hline
\cellcolor{yellow}&\cellcolor{yellow} &\cellcolor{yellow} &\cellcolor{yellow} &\cellcolor{yellow} &\cellcolor{yellow} &\cellcolor{yellow} &  $\cellcolor{yellow}-0.322\pm0.015$ & $\cellcolor{yellow}-0.006\pm0.004$ \\
\multirow{-2}{*}{\cellcolor{yellow}SKAMPLUS} & \multirow{-2}{*}{\cellcolor{yellow}314} & \multirow{-2}{*}{\cellcolor{yellow}197} & \multirow{-2}{*}{\cellcolor{yellow}39} & \multirow{-2}{*}{\cellcolor{yellow}Logarithmic} & \multirow{-2}{*}{\cellcolor{yellow}1.38}& \multirow{-2}{*}{\cellcolor{yellow}31.42} & $\cellcolor{yellow}-0.289\pm0.017$ & $\cellcolor{yellow}0.005\pm0.005$ \\ 
\hline
\end{tabular}}
\end{table}

\begin{figure}
\centering
\includegraphics[scale=0.6]{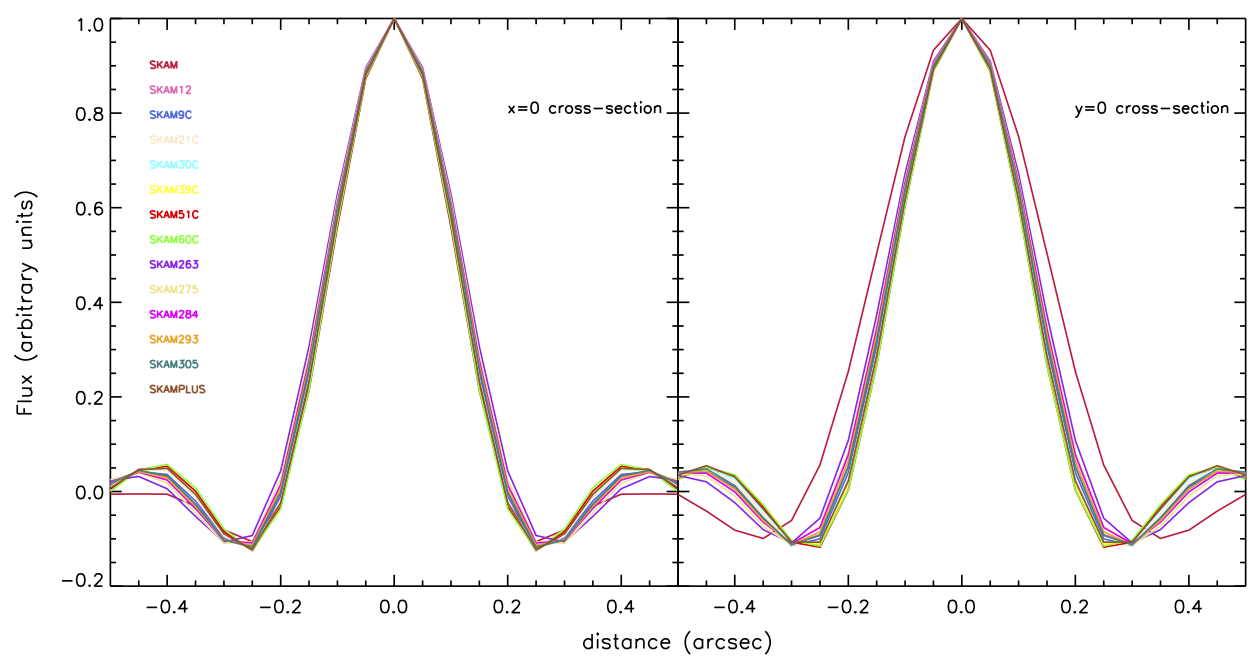} 
\caption{Cross-sections through the PSFs obtained from configurations described in 
\S~3.
\label{fig:skaconfigpsfs} 
} 
\end{figure}

Since these are only modest changes to the configuration (i.e. $\sim20\%$ movement/addition of dishes) we see no significant improvements in performance of the recovered ellipticity distributions. Since the deconvolution of the PSF in known to be one of the major causes of systematic error in shear measurement, in Figure \ref{fig:skaconfigpsfs} we show cross-sections of the PSFs for all the considered configurations. As can be seen, the change in the PSF is small and so reaffirms the result of consistent calibration values in the absence of any other potential causes of noise. 

\subsection{Alternative Configurations}
For weak lensing the main aspect of the baselines configuration is concerned with achieving high sensitivity at scales where we can measure the shapes of sources in the continuum. This translates roughly to having significant sensitivity at scales corresponding to 0.5 - 1 arcseconds. For this reason increasing the the number of antennas in the spiral arms out to 70-80 km is beneficial for weak lensing while the lack of these baselines makes such a survey unfeasible.

To accommodate the other 3 cosmology science cases (cosmology with continuum and HI galaxy surveys and intensity mapping) as well as weak lensing alternative configurations have been proposed such that the $uv$ coverage is as full as possible out to baselines of 70 - 80 km. In order to achieve a smooth transition between the three sections of the array (inner core, outer core and the spiral arms), the core is `puffed' up slightly while the total number of dishes in the (inner+outer) core is preserved. The two proposed configurations are referred to using the following convention SKA1W{\it i}-{\it j}A{\it k}B{\it l}, where {\it i} refers to the number of dishes moved from the outer core to the spiral arms, {\it j} is the number of new dishes added to the arms, {\it k} is the maximum extent of the spiral arms and {\it l} the maximum baseline. The latter two are fixed in both cases to {\it k}$=72$km and {\it l}$=120$km. We shall refer to the two configurations, SKA1W9-0A72B120 and SKA1W9-12A72B120, as A and B respectively.


To assess the capabilities of these alternative configurations we run a new set of simulations at 600 MHz, 800 MHz and 1000 MHz with 1 50 MHz channel observing at declination $\delta=-30^{\circ}$. To compare we also run the same simulation using the Robert Braun SKA1-Mid configuration discussed above, the results are presented in Table \ref{tab:spheconfigs}.

We notice immediately that as we increase the frequency of the observation, the calibration values decrease. This is because the resolution is $\propto \lambda/D$, where $\lambda$ is the wavelength of observation and $D$ the maximum baseline. In this case, our PSF is effectively becoming smaller, and so the higher the frequency, the more accurate our shape measurement. Also, as described in \citet{SPHECONFIGS} these arrays have optimised the distribution of dishes in the spiral arms such that the sensitivity at angular scales of $0.4 - 1$ arc second at 650 MHz can be enhances without significantly compromising the larger scales, so we expect them to perform best at the lower part of the frequency space explored here. 

\begin{figure}
\centering
\includegraphics[scale=0.4]{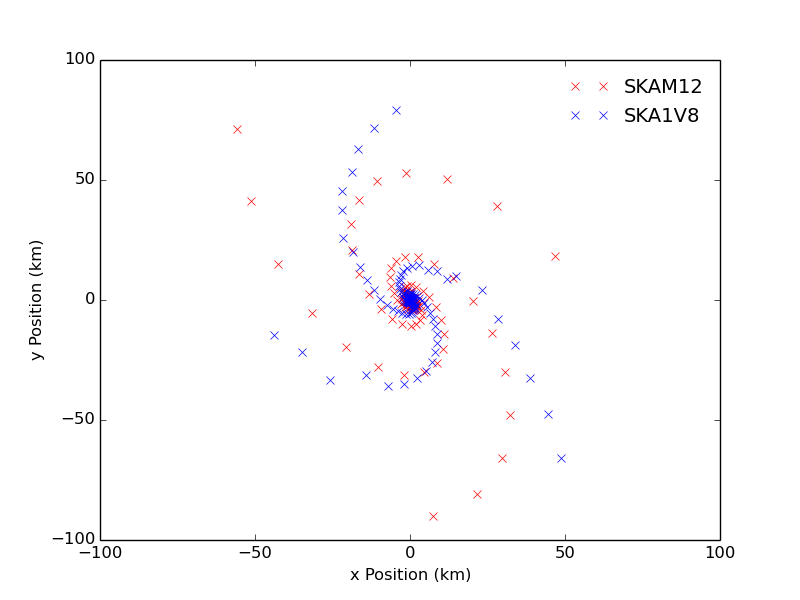}
\caption{SKA1-Mid configurations proposed by Robert Braun (blue) and also one that is smaller in extent and takes into consideration the site geography (red). Results for both these configurations are compared in Table 2.
\label{fig:ref2v8} }
\end{figure}

Also shown in Table \ref{tab:spheconfigs} is the same simulation ran for the so-called SKA1V8 configuration. This is a slightly altered configuration of SKA1-Mid that also has a smaller extent than the two originally proposed configurations (maximum baseline of $\simeq150$km opposed to $\simeq170$km), it also takes into account the geography of the site. This configuration is plotted (along with SKAM12 from above) in Figure \ref{fig:ref2v8}. Encouragingly, even with site topology incorporated this configuration produces similar calibration values to the one originally proposed that did not take this into account, while also bringing down the maximum extent of the spiral arms.


\begin{table}
\caption{Weak lensing simulations results for SKAM12 proposed by Robert Braun and also 2 alternatives (SKA1W9-0A72B120 and SKA1W9-12A72B120) that give a fuller $uv$ coverage to baselines between 70 - 80 km. SKA1V8 is the SKA1-Mid configuration with a smaller extent and with site geography considered. \label{tab:spheconfigs}}
\resizebox{\linewidth}{!}{\begin{tabular}{|c|cc|cc|cc|}
\hline
Array Configuration & \multicolumn{2}{c}{600 MHz} & \multicolumn{2}{c}{800 MHz} & \multicolumn{2}{c}{1000 MHz} \\ & $m_{i}$ & $c_{i}$ & $m_{i}$ & $c_{i}$ & $m_{i}$ & $c_{i}$ \\
\hline
\hline
\multirow{2}{*}{SKA1 REF2 (Robert Braun)} & $-0.560\pm0.039$ & $0.028\pm0.011$ & $-0.508\pm0.031$ & $0.052\pm0.008$ & $-0.424\pm0.021$ & $0.021\pm0.005$ \\ 
& $-0.491\pm0.040$ & $-0.004\pm0.010$ & $-0.434\pm0.033$ & $-0.007\pm0.008$ & $-0.400\pm0.021$ & $-0.001\pm0.005$ \\

\multirow{2}{*}{SKA1W9-0A72B120} & $-0.655\pm0.027$ & $0.033\pm0.007$ & $-0.604\pm0.020$ & $0.034\pm0.005$ & $-0.533\pm0.020$ & $0.016\pm0.005$ \\ 
& $-0.639\pm0.025$ & $-0.006\pm0.006$ & $-0.557\pm0.021$ & $-0.006\pm0.005$ & $-0.530\pm0.021$ & $0.003\pm0.005$ \\

\multirow{2}{*}{SKA1W9-12A72B120} & $-0.582\pm0.043$ & $0.038\pm0.011$ & $-0.563\pm0.020$ & $0.008\pm0.005$ & $-0.530\pm0.018$ & $0.002\pm0.004$ \\ 
& $-0.596\pm0.040$ & $-0.006\pm0.010$ & $-0.532\pm0.020$ & $-0.007\pm0.005$ & $-0.530\pm0.017$ & $-0.008\pm0.004$ \\

\multirow{2}{*}{SKA1V8} & $-0.545\pm0.042$ & $0.026\pm0.011$ & $-0.519\pm0.037$ & $0.071\pm0.010$ & $-0.458\pm0.023$ & $0.052\pm0.006$ \\ 
& $-0.506\pm0.046$ & $0.016\pm0.011$ & $-0.480\pm0.036$ & $-0.003\pm0.009$ & $-0.428\pm0.024$ & $0.004\pm0.020$ \\

\hline
\end{tabular}}
\end{table}

\subsection{SKA Capabilities}
\begin{figure*}
\centering
\includegraphics[scale=0.4]{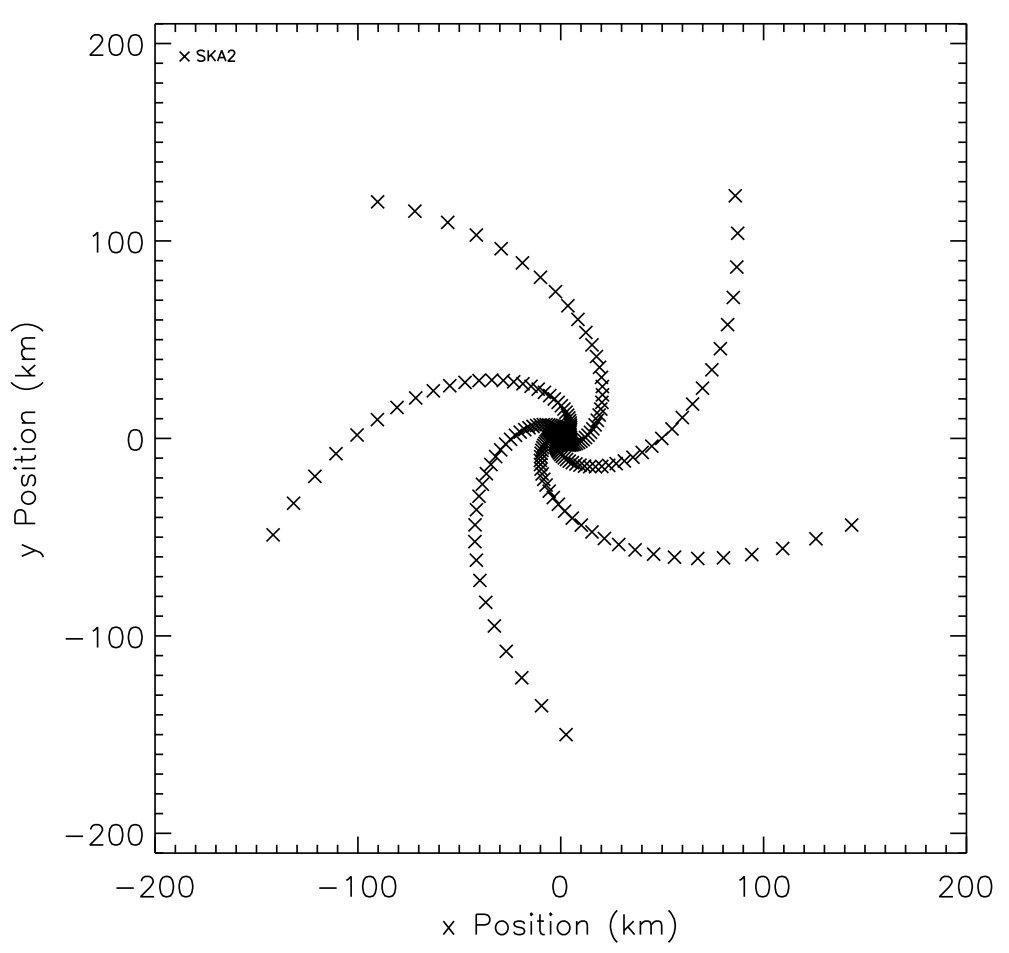} \includegraphics[scale=0.4]{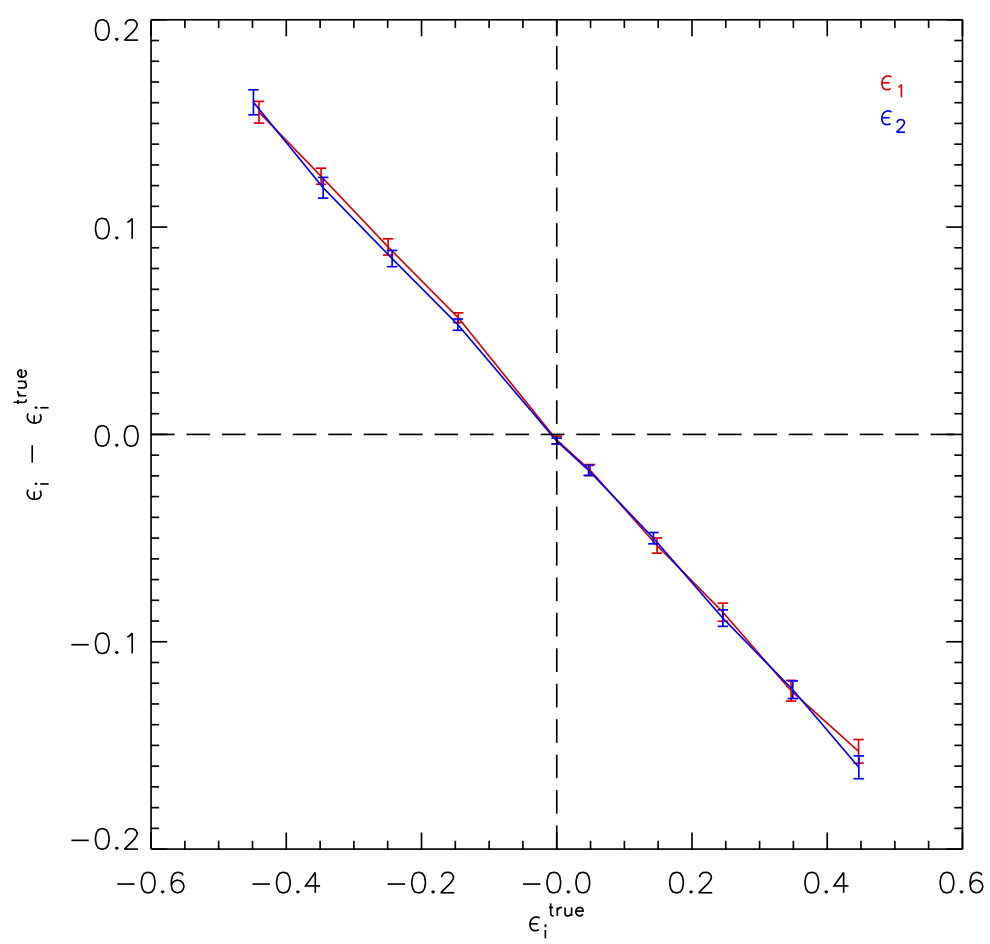}
\caption{{\it Left panel:} A mock SKA configuration with 5 spiral arms extending to 150 km, but no dishes within the inner 1km of the core. {\it Right panel:} Recovered ellipticity distribution for SKA.  \label{fig:ska2config}}
\end{figure*}

In this Section we compute the performance of SKA and compare it to SKA1-Mid. In order to do this we have created a SKA configuration that consists of 5 spiral arms extending out too 150 km (each spiral arm has 50 dishes logarithmically spaced), but we have neglected all the dishes within the 1 km core. This is due to the very large number of baselines involved in a full SKA simulation (and hence massive computation time). Instead, since weak lensing is primarily concerned with the longer baselines anyway, and also at this stage we are only interesting gaining some idea as to what SKA might be able to achieve, we adopt this simplification. The resultant SKA configuration is shown in the left hand panel of Figure \ref{fig:ska2config}.

To compare appropriately to the SKA1-Mid simulations discussed in \S \ref{sec:configs} we again run the SKA simulation with an 8 hour observation at 700MHz and 10 50 MHz channels pointing at declination $\delta=-40^{\circ}$. We have also again corrupted the visibilities by an amount that results in sources being at a SNR $\simeq10$. The received ellipticity distribution is shown in the right hand panel of Figure \ref{fig:ska2config}, the resulting calibration values are:
\begin{eqnarray}
m_{1}&=& -0.357\pm 0.005  \nonumber \\
c_{1} &=& 3\times10^{-4} \pm0.001 \nonumber \\
m_{2} &=& -0.354\pm 0.005 \nonumber \nonumber \\
c_{2} &=& -0.002 \pm 0.001. \nonumber \\
\end{eqnarray}

We see that the multiplicative bias values recovered from SKA seem to be worse than for SKA1-Mid. We note at this stage that this is only a notional SKA configuration that we have simulated and so can not be completely relied upon when comparing to the more sophisticated SKA1-Mid configurations. Since our SKA configuration has some dishes missing, we are invariably missing many short and intermediate length baselines, that also carry shape information of scales that are relevant. We also see that error bars on $m_{i}$ are a factor of 10 smaller than for SKA1-Mid. This is due to the many more sources that reach our final catalogue as SNR $\sim10$ sources, meaning this is a more precise measurement. 

\subsection{Calibration Requirements for SKA}
To provide some context for the obtained calibration values, we calculate the requirements on $m$ and $c$ for stages of SKA1-Mid and SKA and also for comparison, current and future optical surveys such as the Dark Energy Survey \citep{2005astro.ph.10346T}, Euclid and LSST. We adopt the requirements as computed in \citet{2008MNRAS.391..228A}, which are based upon the parameters: sky area, galaxy median redshift and galaxy number density.  The requirements are set such that the statistical error is equal to the systematic error and thus provides an upper limit on the level of bias allowed. 

In addition, we follow the convention of converting the multiplicative and additive biases into a single {\it quality factor} $Q$, computed here as in \citet{2010MNRAS.404..458V} with an assumed rms cosmic shear of $\sigma_{\gamma} = 0.03$. In Table \ref{tab:m_and_c} we show the requirements for notional surveys conducted with an early phase of SKA1-Mid (SKA1-Mid early), SKA1-Mid and SKA, along with corresponding numbers for DES-like and Euclid/LSST-like surveys for comparison. 

The values quoted for the number density and median redshifts are derived for an envisaged 2 year (net) continuum survey over 3 possible survey areas. The specification used are those given in \citet{SKA1IMAGING}, which in turn use the SKA1-Mid baseline design and the SKADS simulations of \citet{2008MNRAS.388.1335W}. The sensitivity levels have been chosen appropriately for weak lensing angular scales of 0.5 arc seconds at Band 2 and the galaxy number densities correspond to $>10\sigma$ detections. SKA1-Mid early is defined to be such that it has 50\% of the sensitivity of SKA1-Mid. 

We note how these requirements are orders of magnitude smaller than those derived in the preceding section. In our simulations we have not attempted to optimise any of the parameters (either in the simulation or the shape measurement analysis) to seek out the smallest calibration values, e.g. we have made no attempt to optimise the imaging of the simulated data by investigating other imaging methods other than CLEAN. The requirements quoted represent the levels that need to be achieved in order for the error budget to be equal between the systematics and statistics. We hope that we can utilise our pipeline further to understand the various systematics and explore different imaging techniques etc. to provide more robust values of for the calibration biases. 

\begin{table}
\centering
\caption{Requirements on multiplicative and additive biases on ellipticity measurement for proposed SKA weak lensing surveys to be dominated by statistical rather than systematics uncertainties, and for DES-like and Euclid/LSST-like for comparison. $Q$ is calculated from $m$ and $c$ as in \citet{2010MNRAS.404..458V}.}
\label{tab:m_and_c}
\begin{tabular}{lcccccc}
\hline
Experiment & $A_{sky}$ & $n_{gal}$ & $z_{m}$ & $m<$ & $c<$ & $Q>$ \\
\hline
\hline
DES-like & 5000 & 12 & 0.8 & 0.004 & 0.0006 & 260 \\
Euclid/LSST-like & 20000 & 35 & 0.9 & 0.001 & 0.0005 & 990\\
\hline
SKA1-Mid early & 1000  & 3.0 & 1.0 & 0.014   & 0.0012  & 62 \\
SKA1-Mid early & 5000  & 1.2   & 0.8 & 0.012  & 0.0011  & 79\\
SKA1-Mid early & 30940 & 0.35   & 0.5 & 0.011  & 0.0011 & 80\\
\hline
SKA1-Mid       & 1000  & 6.1   & 1.2 & 0.0090   & 0.00095  & 103\\
SKA1-Mid       & 5000  & 2.7   & 1.0 & 0.0067  & 0.00082 & 140\\
SKA1-Mid       & 30940 & 0.9 & 0.7 & 0.0058  & 0.00076 & 164\\
\hline
SKA       & 1000  & 37    & 1.6 & 0.0031  & 0.00055 & 318\\
SKA       & 5000  & 23    & 1.4 & 0.0019  & 0.00043 & 523\\
SKA       & 30940 & 10    & 1.3 & 0.0012 & 0.00035 & 825\\
\hline
\end{tabular}
\end{table}

\section{Shear Measurement Techniques}
As discussed above, the signal in weak gravitational lensing is the small shearing of galaxy images by foreground matter. The smallness of this shearing (typically of order $1\%$) and its sensitivity to change in cosmological parameters (typically of order $0.01\%$ for a $1\%$ change in the dark energy equation of state $w$) means that any effect which is capable of biasing results must be carefully controlled. One place in which such a bias may enter is in the translation from real, noisy data to a map of shear measurements across the field-of-view. Typically this is done by measuring the ellipticity of galaxies identified in the data, which is changed by shear. The preponderance of optical data of the quality necessary for weak lensing has led to the development of a large number of different techniques for performing this shape measurement process which take imageplane data as their inputs. Among the first derived (and subsequently most widely used) are methods which use weighted quadropole moment of combined galaxy-PSF images to measure ellipticities directly in a non-parametric way (KSB \cite{1995ApJ...449..460K}; KSB+ \cite{1998ApJ...504..636H}; Re-Gaussianization \cite{2003MNRAS.343..459H}). Another popular approach is to assert that the galaxy images may be modelled with some analytic brightness distribution (such as a Gaussian or Sersic profile) and find the best fitting parameters, including ellipticity parameters, for each source (IM3SHAPE \cite{2013MNRAS.434.1604Z}, lensfit \cite{2007MNRAS.382..315M}). 

In the radio regime the approach which has found most application is that of shapelets, which reconstructs the data using an orthonormal set of basis functions. How these basis functions transform with shear is known, meaning the best-fitting coefficients for an image may be used to form an unbiased estimator for the shearing it has undergone through comparison with coefficents from some `unlensed' sample. Shapelets also have the advantage of having similarly simple and analytic Fourier transformations which also remain localised, facilitating their use in directly modelling visibility data rather than reconstructed images.

As demonstrated above, using simulations with known ellipticity distributions provides a way of probing different aspects of the weak lensing pipeline. Most notably in the optical community simulations have been used for testing shape measurement techniques.
Over the previous 10 years, the Shear TEsting Programme (STEP) and GRavitational lEnsing Accuracy Testing (GREAT) (see \cite{2014ApJS..212....5M} and references therein for a brief history) initiatives have simulated large optical weak lensing datasets and invited participants to (blindly) measure the shear in those images. This has allowed relative proficiency of different shear measurement methods and how they react to changes in data parameters, such as source size, S/N and simulated galaxy model complexity, to be quantified. These investigations have given insight into the behaviour of shear measurement algorithms in optical data, but we may expect the challenges of radio data to be significantly different. 

The production of images from interferometer data via deconvolution using algorithms such as CLEAN is a highly non-linear process and has the potential to produce spurious cosmic shear signals in addition to (and in convolution with) any introduced by a particular shape measurement algorithm. The noise in radio images is highly correlated (though optical methods have experience with this in dealing with multi-epoch data). The ability of presently available techniques to deal with these challenges is currently unclear, motivating a systematic evaluation along the model of STEP and GREAT programmes. A direct follow-on from these efforts may evaluate the ability of optical shape measurement algorithms to be extended to radio images, but a worthwhile comparison may also be done of the image reconstruction algorithms themselves, either in separation or in conjunction with shear estimation; in addition to long-established algorithms such as CLEAN and Maximum Entropy approaches, several new methods are under active research (e.g. \cite{2014MNRAS.438..768S}, \cite{2014MNRAS.439.3591C} and references therein).

If image-reconstruction algorithms are unable to perform to the low level of systematics necessary for weak lensing science, it may be necessary to perform shear measurement directly using the visibilities produced by the telescope. Indeed, the only current detection of a shear signal in radio sources was performed in the visibility plane with shapelets. 
Notionally, being able to avoid the deconvolution process necessary for imaging is an important strength of visibility-plane shear measurement methods. The deterministic nature of instrumental effects in radio astronomy is often touted as a key advantage of radio weak lensing and it manifests here through the ability to forward convolve sky models in a well-defined way when searching for the best fitting parameters. However, the computational challenges of such a procedure are potentially great. Information from each individual source is no longer localised in the visibility plane, meaning all sources must be fitted simultaneously, rather than simply taking cut-outs of images around a single source as is done in image-plane methods. Starting with a naive model-fitting approach, if we attempt to fit ~5 parameters per source (as is typically done in mainstream optical methods) for large numbers of sources over (extremely) large numbers of visibilities, this quickly becomes computationally unfeasible. There are various potential ways of alleviating these problems: data volumes could be reduced with averaging of visibilities on a grid, the number of necessary simultaneous fits could be reduced by averaging around a single source at the phase centre or by employing methods such as visibility-stacking \citep{Lindroosinprep} and we may also expect other, novel techniques to be developed. In addition to the computational problems involved in the analysis, the sheer volume of storage necessary to maintain access to unaveraged visibilities from the SKA may also be prohibitive.

\subsection{radioGREAT}
In order to investigate the issues discussed above, the radioGREAT\footnote{http://radiogreat.jb.man.ac.uk} programme has been initiated, with three key goals:
\begin{itemize}
	\item What are the requirements on shape measurement for cosmology with weak gravitational lensing in the radio band?
	\item Can we make images of the necessary fidelity to measure shapes of radio star-forming galaxies to the level of these requirements?
	\item Can we measure shapes of radio star forming galaxies to the level of these requirements whilst leaving our data in the visibility plane?
\end{itemize}
For an initial challenge, the task should be kept as simple as possible, with complications introduced individually in order to evaluate their effect on shape measurement. We may expect to follow closely the structure of the GREAT08 optical/NIR challenge, with a fiducial branch consisting of non-overlapping, simple Sersic galaxy models with identical, high signal-to-noise values ($>\sim100$) and constant radii. Other branches may then consist of datasets containing individual modifications, to systematically evaluate the effect of e.g. altering galaxy size, reducing signal-to-noise, altering dynamic range and altering bandwidth and time smearing.

\subsection{Shapelets in Real and Fourier Space}
\label{sec:shapelets}
\noindent In contrast to traditional telescopes interferometers do not provide a direct image of the observed sky, but instead measure its Fourier transform at a finite number of $uv$ points that correspond to each antenna pair in the array. The real space image must then be reconstructed from this discretely sampled visibility data, while also deconvolving the effective beam that arises from the finite sampling. Several methods exist to perform this task, e.g. CLEAN, MEM. These methods are well tested and are appropriate for various applications; however, these methods are non-linear and do not necessarily converge in a well defined manner. 

The Hermite polynomials that form the shapelet basis set have some remarkable properties which greatly facilitate the modelling of the source shapes. Of particular interest here is the property that they are invariant (up to a rescaling) under of Fourier transform and thus are naturally suited for interferometric imaging. In optical surveys, shapelets have been little utilised after some conceptual concerns were raised with the approach (of shapelets like methods), however, similar conceptual problems were also raised with the KSB method although this method continued to be popular in the literature, see \citet{2011MNRAS.412.1552M} for more details.

In the real space (or image plane) application of the shapelets technique the surface brightness $f({\bf x})$ of an object is decomposed as
\begin{equation}
f({\bf x})=\sum_{{\bf n}}f_{{\bf n}}B_{{\bf n}}({\bf x};\beta);
\end{equation}
where
\begin{equation}
B_{{\bf n}}({\bf x};\beta)=\frac{H_{n_{1}}(\beta^{-1}x_{1})H_{n_{2}}(\beta^{-1}x_{2})e^{-\frac{|x|^{2}}{2\beta^{2}}}}{\left[2^{(n_{1}+n_{2})}\pi\beta^{2}n_{1}!n_{2}!\right]^{\frac{1}{2}}}
\end{equation}
are the two-dimensional orthonormal basis functions with a characteristic scale $\beta$, $H_{m}(\eta)$ is the Hermite polynomial of order $m$, ${\bf x}=(x_{1},x_{2})$ and ${\bf n}=(n_{1},n_{2})$. The basis functions are complete and if $\beta$ and ${\bf x}$ are chosen to be close to the size and location of the galaxy, then the expansion will yield a quick convergence.


The Fourier transform of an objects intensity can be written as 
\begin{equation}
\tilde{f}({\bf k})=(2\pi)^{-\frac{1}{2}}\int_{-\infty}^{\infty}f({\bf x})e^{i{\bf k}\cdot{\bf x}}d^{2}x 
\end{equation}
and decomposed as 
\begin{equation}
\tilde{f}({\bf k})=\sum_{{\bf n}}f_{{\bf n}}\tilde{B}_{{\bf n}}({\bf k};\beta), 
\end{equation}
where $\tilde{B}_{{\bf n}}({\bf k};\beta)$ obey the dual property
\begin{equation}
\tilde{B}_{{\bf n}}({\bf k};\beta)=i^{(n_{1}+n_{2})}B_{{\bf n}}({\bf k};\beta).
\end{equation}
The invariance (unto a rescaling) under Fourier transform makes this basis set a natural choice for interferometric imaging.

As mentioned above, an interferometer correlates the signals measured by antenna pairs into a complex visibility. Each of these visibilities occupies a point on the $uv$ plane which corresponds to the projected baseline formed between the antenna pair. In practice, the visibilities are not exactly a two-dimensional Fourier transform of the sky brightness. The visibility measured for an antenna pair $(i,j)$ at a time and frequency of $t$ and $\nu$ respectively is given by
\begin{equation}
V_{ij}(\nu,t)=\int\frac{A({\bf l},\nu)f({\bf l},\nu,t)}{\sqrt{1-|{\bf l}|^{2}}}e^{-2\pi i\left[u\ell+vm+w(\sqrt{1-|{\bf l}|^{2}}-1)\right]},
\end{equation}
where $f({\bf l},\nu,t)$ is the surface brightness of the sky at location ${\bf l}=(\ell,m)$ with respect to the phase centre and $A({\bf \ell},\nu)$ is the frequency dependent primary beam. The $(u,v,w)$ coordinates are given by 
\begin{equation}
\begin{pmatrix} u \\ v \\ w  \end{pmatrix} =
\begin{pmatrix} \sin H_{0} & \cos H_{0} & 0 \\
			-\sin\delta_{0}\cos H_{0} & \sin\delta_{0}\sin H_{0} & \cos\delta_{0} \\
			\cos\delta_{0}\cos H_{0} & -\cos\delta_{0}\sin H_{0} & \sin\delta_{0}
\end{pmatrix}
\frac{1}{\lambda}
\begin{pmatrix} L_{x} \\ L_{y} \\ L_{z} \end{pmatrix},
\end{equation}
where $\lambda$ is the wavelength of the observation, $H_0$ is the hour angle and $\delta_{0}$ the declination. $L_{x,y,z}$ are the coordinate differences between the two antennas measured in a fixed-Earth coordinate system in which the sky rotates about the $\hat{L}_{z}$ axis. We also note that locus that is traced by the $(u,v,w)$ coordinates also define the synthesised beam pattern (or PSF). Since these coordinates are entirely determined from the antenna and source positions and the time and frequency of the observation, the synthesised beam is entirely known for interferometers. Only for observations at the zenith $(z\simeq 0)$, in the absence of a primary beam $(A\simeq1)$ and for small displacements from the phase center $(\ell,m << 1)$ does the relationship between the measured visibilities and the desired sky brightness reduce to an extact Fourier transform.

Since visibility datasets can be large $(>10^{5})$, directly fitting the shape parameters to all the $uv$ data can be very computationally expensive. Instead one can apply a binning scheme to reduce the size of the data but without losing any information. In the $uv$ plane we can grid the data using a cell size $\Delta u = 0.5\Delta\ell^{-1}$, and average the data in each cell and similarly for $\Delta v$. $\Delta\ell$ is chosen to be one half of the intended field-of-view and the 0.5 factor accounts for the Nyquist frequency. This choice of $\Delta u,v$ is designed to minimise the number of cells but also to avoid smearing at large scales which mimic primary beam attenuation.

In our implementation of this technique we model the intensity $f_{s}(\ell,m)$ of each source $s$ as a sum of the shapelet basis functions $B_{\bf n}({\bf l-l}_{s}$;$\beta_{s})$, centred on the source centroid ${\bf l}_{s}(\ell_{s},m_{s})$ with scale $\beta_{s}$, by estimating the shapelet coefficients $f_{{\bf n}s}$ of the sources given the binned visibility data described above. In principle, a fully sampled $uv$ plane provides all the shape information about the sources and a linear decomposition similar to the image domain described above could be performed. However, given the finite sampling in the $uv$ plane this is not possible here. To alleviate this problem we make a linear fit to the $uv$ plane with the shapelet coefficients as free parameters. 
\subsection{$uv$ Plane shapelet Fitting}
\begin{figure}
\centering
\includegraphics[scale=0.5]{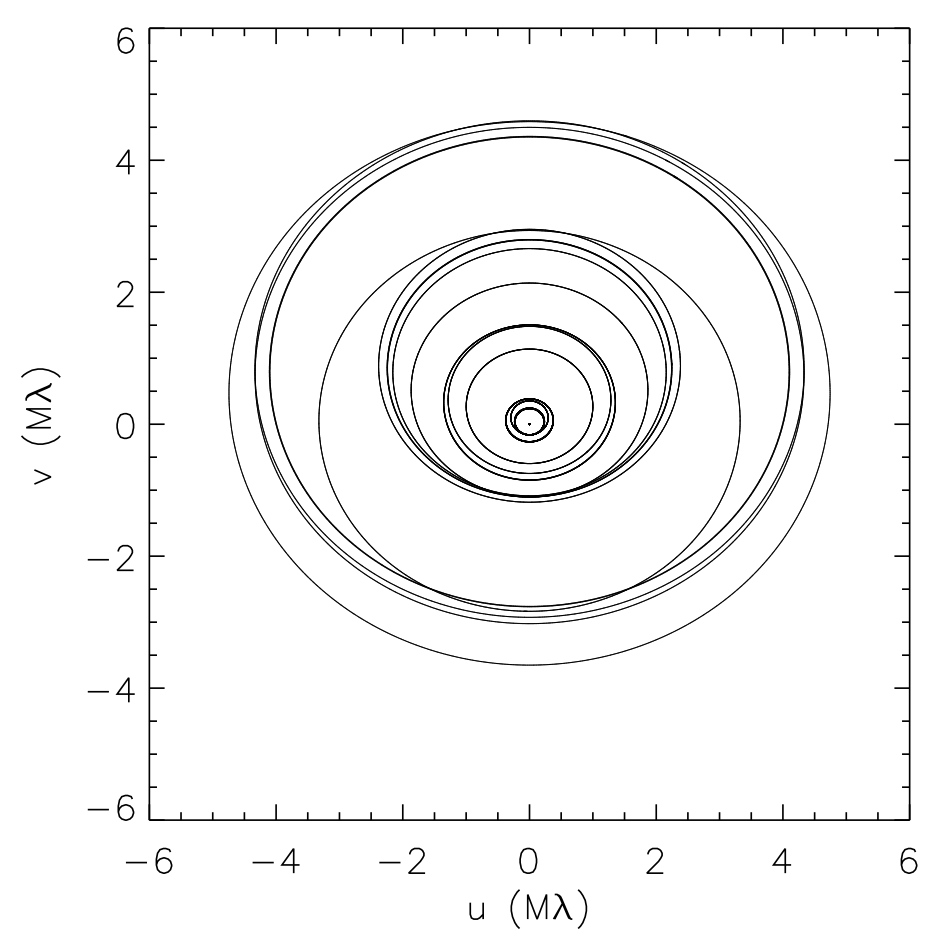}
\caption{$uv$ coverage of the eMERLIN experiment used to demonstrate shape measurement using shapelet fitting in the $uv$ plane. \label{fig:emergeuv}}
\end{figure}

To illustrate the technique we have performed simple simulations. We have taken a sample of the source models used in the studies discussed previously and made very simple sky patches containing a grid of $10\times10$ sources. In order to keep the simulations small (i.e. the number of visibilities) we have adopted the eMERLIN array for this experiment. The observation configuration is as follows: the fields are all observed at $\delta=+60^{\circ}$, at 1.4 GHz with a single 125 kHz channel, and for 24 hours and 20 second integration. The resulting $uv$ coverage on which the following results are based is shown in Figure \ref{fig:emergeuv}.
We produce $\sim50$ tiles that have a grid of 100 sources each, each tile consists of sources with the same $n_{max}$, the range of $n_{max}$ is $0-20$. We simulate the visibilities for each of these sources, bin and fit for the shapelet coefficients as described above. As in the previous Sections we then compare the input and output ellipticities. Our current implementation of this technique fits all source with the same number of coefficients (i.e. for the same $n_{max}$), hence our motivation to tile sources according to $n_{max}$. In practice, we would want to fit all sources based on some guess or estimate for the what the $n_{max}$ for each source in our field should be. Likewise, we currently input the source positions and $\beta$ parameters into the code when we do the shape fitting. We are working towards an implementation where the $n_{max}$ and $\beta$ parameters are also fir for. 

\begin{figure*}
\centering
\includegraphics[scale=0.5]{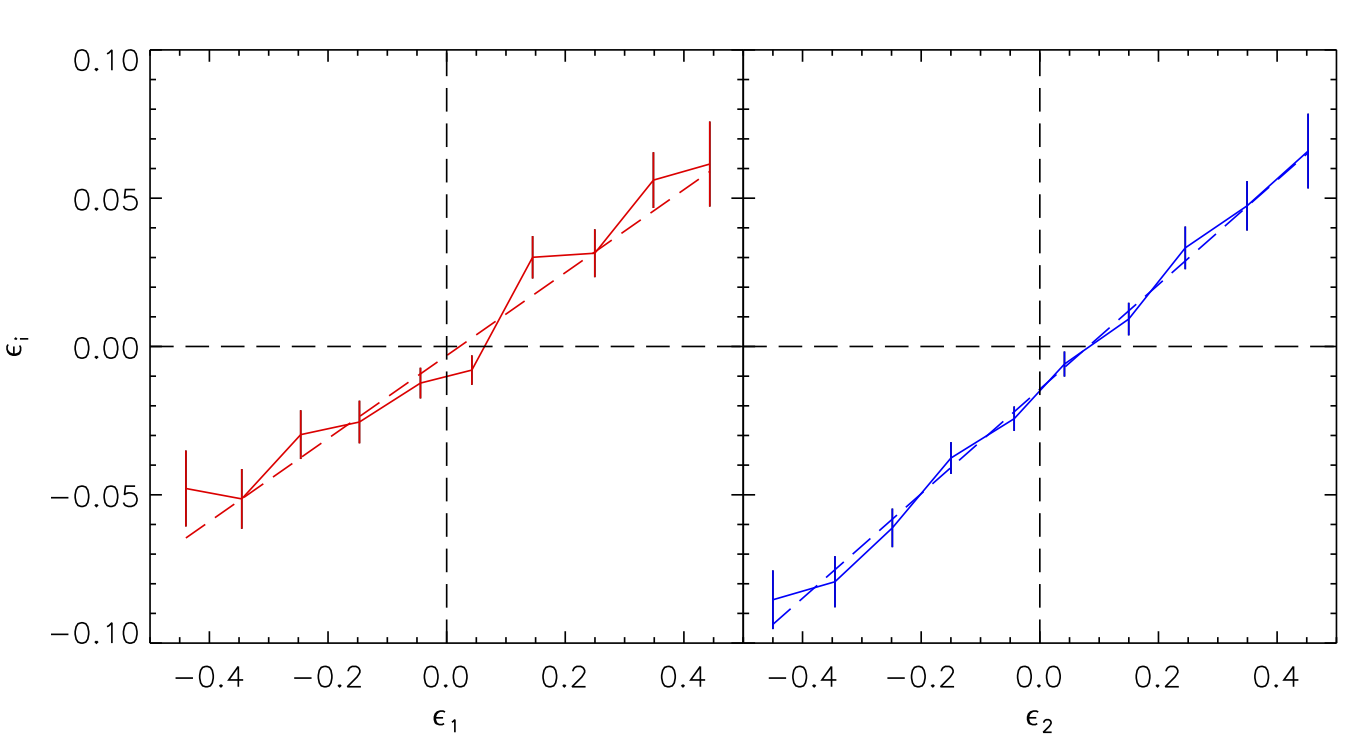}
\caption{Ellipticity comparison based on shapelet fitting in the $uv$ plane. The solid curves are show the binned data and the dashed curves are the best linear fits. The red lines are for $\epsilon_{1}$ and the blue for $\epsilon_{2}$. \label{fig:uvfit4} }
\end{figure*}

As an initial study, in Figure \ref{fig:uvfit4} we show the comparison of input and output ellipticities having fit all tiles with an $n_{max}=4$, this is the lowest $n_{max}$ we can chose for our adopted ellipticity estimator. 
We see that the bias we achieve in this (simplified) scenario is much less than anything we achieved in the image plane analyses of the previous Sections, the calculated bias values are
\begin{eqnarray}
m_{1}&=&0.140\pm0.012 \\ \nonumber
c_{1}&=&-0.003\pm0.002 \\ \nonumber
m_{2}&=&0.176\pm0.010 \\ \nonumber
c_{2}&=&-0.014\pm0.002 
\end{eqnarray}
Once again we find very little additive bias while a much smaller multiplicative bias than we have encountered elsewhere in this chapter. Although this in not a meaningful comparison as the input images and simulations are fundamentally different in both cases. We aim to expand these $uv$ plane simulations considerably in order to make much better comparison between the performance of shapelet fitting in the $uv$ and image planes. Ultimately we hope to extend this to real data that has already been analysed in the the image plane in \citet{2010MNRAS.401.2572P}. 
\section{Discussion/Conclusion}
Over recent times weak lensing has emerged as a very promising tool for cosmology. Making weak lensing measurements requires the measurement of very many galaxy shapes to high precision, while also carefully controlling systematics. Due to the involved nature of making radio images making use of simulated data provides a important way to probe such systematic effects in the system while also being key to testing current and new techniques. 

In this chapter we have provided details of the ongoing efforts carried out to address the feasibility of weak lensing experiments with SKA1-Mid and SKA in the context of practical shape measurement. We have implemented a pipeline that allows us to explore many aspects of the data gathering and reduction processes, allowing us to asses if these are able to meet the high demands on data quality that weak lensing measurements require. 

We have explored how small changes to the proposed SKA1-Mid array configuration has little impact on ones ability to accurately measure the shapes of galaxies. We have introduced alternative configurations that are smaller in overall extent that give similar performance to those originally proposed and which have a better survey speed. Also, demonstrated was the performance of SKA1-Mid with the site geography taken into consideration and this provides encouraging results. 

Using a notional SKA configuration we find slightly diminished calibration biases in comparison to SKA1-Mid but this result requires further investigation as it is unlikely this can truly be the case. A more complete configuration without missing baselines should give a much clearer idea of how SKA will perform in comparison on SKA1-Mid. 

In our present studies we have not tried to optimise the simulations or shape measurement analysis in order to achieve the optimal calibration biases. However, we did present the requirements that will be required for constraining cosmology with real SKA surveys. We can make use of the pipeline that we have developed to determine where these biases originate. 

We introduced the forthcoming radioGREAT challenge that will investigate current and new methods of shape measurement and their applicability to radio data. Starting with simple simulations we hope to emulate the successes of the STEP and GREAT challenges in the optical to address the difficulties that radio data gathering and reduction will most likely pose. We also demonstrate the shapelet shape measurement technique as applied directly to simple visibility data. In due course, we hope to further develop these simulations to more realistic scenarios and to properly compare the technique applied in both image and visibility plane, before eventually applying to real data. 


\section*{Acknowledgements}
PP is funded by a SKA South Africa Postdoctoral Fellowship. IH and MLB are supported by an ERC Starting Grant (Grant no. 280127). MLB is a STFC Advanced/Halliday fellowship. SM acknowledges financial support from the National Research Foundation of South Africa. OS research is supported by the South African Research Chairs Initiative of the Department of Science and Technology and National Research Foundation. FBA acknowledges the support of the Royal society via an RSURF. MJJ acknowledges support by the South African Square Kilometre Array Project and the South African National Research Foundation. DB is supported by UK Science and Technology Facilities Council, grant ST/K00090X/1.
\bibliographystyle{apj}
\bibliography{../../mybib.bib}

\end{document}